# Correlation between defect and magnetism of $Ar^{9+}$ implanted and un-implanted $Zn_{0.95}Mn_{0.05}O$ thin films suitable for electronic application

S. K. Neogi[a], N. Midya[a], P. Pramanik[b], A. Banerjee[a,c], A. Bhattacharya[b,c], G. S. Taki[d], J.B.M. Krishna[e] S. Bandyopadhyay[a,c*]

[a] Department of Physics, University of Calcutta, 92 APC Road, Kolkata:700009, India
[b] Institute of RadioPhysics and Electronics, University of Calcutta, 92 APC Road, Kolkata:700009, India
[c] CRNN, University of Calcutta, JB Block, SectorIII, SaltLake, Kolkata:700098, India
[d] Variable Energy Cyclotron Centre, 1/AF, Salt Lake, Kolkata: 700064, India
[e] UGC DAE CSR, Kolkata Centre, LB 8, Sector III, Salt Lake, Kolkata: 700098, India

**Abstract**

Sol-gel derived thin films of $Zn_{0.95}Mn_{0.05}O$ have been implanted with $Ar^{9+}$ ions with doses viz. $5\times10^{14}$ ions/cm$^2$ (low), $1\times10^{15}$ ions/cm$^2$ (intermediate) and $1\times10^{16}$ ions/cm$^2$ (high). Structural, morphological, optical and magnetic properties of the films have been investigated. Structural study confirmed single phase, wurtzite structure of the films. The absence of impurity phase has been confirmed from several measurements. Ion implantation induces a large concentration of point defects into the films as identified from optical study. All films exhibit well above room temperature (RT) intrinsic ferromagnetism (FM) as evidenced from field and temperature dependent magnetization measurements. The magnetization attains the maximum value for high dose of $Ar^{9+}$ ion implanted film. It shows RT saturation magnetization ($M_S$) value of 0.69emu/gm. The observed FM has been correlated with proportion of intrinsic defects, such as, zinc and oxygen vacancies and the values of $M_S$. Defect induced formation of bound magnetic polaron actually controls the FM. The utility of these films in transparent spin electronic device has also been exhibited.

*Corresponding author: sbaphy@caluniv.ac.in

# I. INTRODUCTION

Dilute magnetic semiconductors (DMS) have attracted much research interest because of their potential applications in the field of spintronics. The euphoria started following the prediction of RT ferromagnetism (FM) in Mn-doped ZnO by Dietl *et al.*[1] The FM between the localized moments of the Mn atoms is mediated by free holes in the material.[1] In ZnO, the exchange interaction for Mn is fundamentally different from the other transition metal ions. Mn *d*-states lie in the valence band while the other transition metals *d*-state introduces states within the gap. Formation of *p*-type ZnO with high carrier density is really a difficult thing. An alternative model shows FM could originate from carriers (holes) that are present in the material, but localized at $Mn^{2+}$ impurity.[2] The bound magnetic polaron (BMP) model[2] shows exchange interaction between $Mn^{2+}$ ion and localized holes that are near to the Mn ion. The size of BMP grows until its radius overlaps that of neighbouring BMP. With emergence of lot of contradictory experimental results in ZnO based DMS system particularly for Mn doping no strong physical understanding has been developed. Recently the clouds of confusion and contradiction hovering over Mn doped ZnO based DMS seems to be dissolved substantially. This is because of emergence of a general consensus that defects play a crucial role in controlling the magnetic properties of such systems.

In reality, generation of ZnO based DMS system is possible if the observed FM is defect mediated. Various types of defects such as zinc vacancy ($V_{Zn}$), oxygen vacancy ($V_O$), zinc interstitial ($I_{Zn}$) may be responsible for FM in Mn doped ZnO system. Liu *et al.* [3] demonstrated that $V_O$, $V_{Zn}$ and $I_{Zn}$ have carried magnetic moment of 0.98 $\mu_B$, 1.998 $\mu_B$, and 2.00 $\mu_B$ respectively. It is indicative of different contributions to FM from different defect species. Among all these defect species the role of $V_{Zn}$ seems to be most favorable for mediating ferromagnetic interaction in $Zn_{1-x}Mn_xO$ system.[4-8] Theoretical calculation on electronic structure and magnetic interaction between Mn ions in the model structure of $Mn_{Zn}$+$V_{Zn}$ shows stability of ferromagnetic states against antiferromagnetic states.[5] Experimentally it has been demonstrated that cation vacancies ($V_{Zn}$) play the determining role in mediating FM for Mn doped ZnO.[6] Also Xu *et al.*[8] experimentally demonstrated that FM in $Zn_{1-x}Mn_xO$ compounds originates due to alignment of magnetic moments mediated by some acceptor defects such as ($V_{Zn}$). Actually a delicate balance of different defect species and attainment of optimum defect concentration along with substitution of Mn at the Zn site is responsible for FM.

In one of our work, [6] it was established that Mn doped ZnO bulk sample synthesized by solid state reaction method is intrinsically ferromagnetic; and the observed FM is defect mediated. We had also studied the effect of $Li^{3+}$ ion irradiation on different physical properties of Mn doped ZnO powder samples previously.[9-10] Defect mediated modification of physical (obviously including magnetic) properties were occurred there.[10] However for the same Mn doped ZnO bulk sample derived from sol-gel route any ferromagnetic ordering [11-12] hasn't been found. So far as sol-gel derived Mn doped ZnO thin films are concerned there are very few reports of achievement of FM.[13] In this article strong ferromagnetism has been observed in sol-gel derived $Zn_{0.95}Mn_{0.05}O$ films up to 325K. The observation of FM at 325K seems to be quite interesting so far as utilization of DMS in spintronic devices are concerned. All electronic components are heated during its operation. Specific cooling arrangements are being made so that temperature of these components remains within control. A Mn doped ZnO thin film operating as DMS in a spintronic device when enable to retain its ferromagnetic properties well above RT can enhance its utility in manifold. Precisely no cooling arrangements will be required at least up to RT. Further observation of FM well above RT also assures the unhindered and flawless operation of this particular DMS system at least up to RT. Another point which is noteworthy is that moderately high temperature ($500^oC$) annealing during growth of these films. Generally there is a prevailing notion that heat treatment temperature particularly in chemical route of synthesis is to be fixed at temperature as low as possible to achieve FM.[14] High temperature (above $400^oC$) heat treatment may kill the inherent FM prevailing in the system. But in this article it has been shown that $500^oC$ annealed $Zn_{0.95}Mn_{0.05}O$ films retain strong FM. If a system can retain its FM, high temperature annealing is always preferable particularly for chemical method of synthesis. The advantage of high temperature annealing in chemical method is twofold. Firstly, polycrystalline structural formation in this type of films is surely better in case of high temperature heat treatment. The other advantage is that removal of organic residue from the surface of film can be done more effectively.

In the present work, the magnetic properties of polycrystalline $Zn_{0.95}Mn_{0.05}O$ films have been analyzed. Low energy (81 KeV) $Ar^{9+}$ ion implantation has been performed to generate defects in Mn doped ZnO films in a controlled fashion. Ion implanted films exhibit degradation of magnetic properties with low dose of implantation. With increasing dose of implantation the trend gradually changed. Ultimately for the film implanted with high dose the magnetic property seems to be the best. The FM has been interpreted from BMP model. However, instead of p-type dopants a particular point defect (Zn vacancy) establishes magnetic exchange with $Mn^{2+}$ ions. The observed trend of FM is highly correlated with formation of Zn and oxygen vacancy type defects inside the system. Overall the idea of defect mediated

FM in ZnO based DMS system seems to be strengthened from this work. The utility of these films in future electronic devices has also been exhibited.

## II. EXPERIMENTAL DETAILS

$Zn_{0.95}Mn_{0.05}O$ thin films were synthesized by sol-gel spin coating technique. Stoichiometric amount of zinc acetate dihydrate $[Zn(CH_3COO)_2,2H_2O]$ and manganese acetate dihydrate $[Mn(CH_3COO)_2,2H_2O]$ was added to a solution containing 2-propanol and diethanolamine (DEA). DEA was used as the sol stabilizer. The precursor solutions thus obtained were used for spin coating on glass substrates to prepare the films. The spinning rate and period were optimized to 2000 rpm and 35 sec, respectively. After coating, the films were dried at $300^0$C for 15 minutes in a furnace to evaporate the solvent. It was followed by heat treatment at moderately high temperature $500^0$C for 30 minutes to get good quality films. The process of coating and subsequent annealing were repeated for five times to obtain the desired film thickness.

The films were implanted with 81 KeV $Ar^{9+}$ ions with doses viz. $5 \times 10^{14}$ ions/cm$^2$, $1 \times 10^{15}$ ions/cm$^2$ and $1 \times 10^{16}$ ions/cm$^2$ at RT. The area of all the films was maintained at (1cm x 1cm). Beam dimension was also maintained at (1cm x 1cm). Rutherford Back Scattering (RBS) measurement has been performed for estimation of the thickness of the film.

Structural, morphological, optical and magnetic properties of un-implanted and implanted 5 at% Mn doped ZnO films were examined. Structural properties were investigated by powder X-ray Diffraction (XRD) technique in glancing angle mode. XRD patterns of the synthesized films were recorded with $CuK_\alpha$ radiation using an automatic powder diffractometer (Make-Philips, Model: PW1830), equipped with ($\theta$-$2\theta$) geometry. The surface morphology of the films was investigated through atomic force microscope (AFM) and magnetic force microscope (MFM). Optical properties were studied by UV-visible and photoluminescence spectroscopy. UV-Visible spectra were recorded using a spectrophotometer (Perkin Elmer; Model: Lambda 35) in the wavelength range (200-1100 nm). The Spectra were recorded by taking a similar glass as the reference and hence transmission due to films only was obtained. Photoluminescence spectra of the films at RT were recorded; the excitation source was the 325-nm line of He-Cd laser with an output power (~10 mW) and monochromator fitted with Photomultiplier tube. Magnetic measurements were performed using superconducting quantum interference device (SQUID) vibration sample magnetometer (VSM) (SQUID VSM, Quantum Design). The field dependent magnetization (M-H) measurement was performed at 300K, 250K and 200K for all of the films. M-H measurement for un-implanted film has been made well above RT (325K) also. The temperature dependent magnetization (M-T) data was recorded in zero field cooled (ZFC) and field cooled (FC) mode at a constant field strength of (150 Oe) in the temperature range 15K to 300K.

## III. RESULT AND DISCUSSIONS

The thickness of the films measured accurately by RBS measurement. The profile of $Zn_{0.95}Mn_{0.05}O$ film emerging from RBS measurement has been shown in figure 1. The film thickness has been estimated by analyzing the RBS spectra. It has been found to be (500 ± 10) nm.[13]

Ion implantation by low energy (81 KeV) $Ar^{9+}$ ion beam with different doses has been performed on $Zn_{0.95}Mn_{0.05}O$ films. Solids under ion bombardment experience complex dynamic annealing processes due to rapid collisions of the target atoms with the incoming ion beam.[15-16] When energetic ion penetrates inside the material, it loses energy mainly by two nearly independent processes: (i) nuclear energy loss represented as ($S_n$) (elastic process), which dominates at an energy of about 1 KeV/amu; and (ii) electronic energy loss represented as ($S_e$) (inelastic process), which dominates at an energy of about 1 MeV/amu or more.[16] The relative contribution of ($S_e$) and ($S_n$) depends on the projectile mass velocity, charge state and the target materials. Electronic energy loss ($S_e$) is mainly responsible for exciting the electrons of the target atoms. Energy loss due to elastic collision ($S_n$) is mainly responsible for knocking out the target atoms hence responsible for the production of large concentration of point defects, vacancy/vacancy cluster etc inside the system.[15-17] For low energy implantation (~ KeV range) nuclear energy loss is dominated over electronic energy loss.[17] The energy losses of $Ar^{9+}$ ions, ($S_e$ and $S_n$) have been estimated by using simulation software: stopping power and ranges of ion in matter (SRIM).[18] For SRIM calculation we have used 4gm/cm$^3$ as the density of the $Zn_{0.95}Mn_{0.05}O$ samples.[16] The displacement threshold energy has been taken to be 18.5 eV and 41.4 eV for Zn and O atom in ZnO lattice[16] respectively. The result of the simulation is shown in figure 2. It has been observed that $S_n$ largely predominates over $S_e$. We obtained nearly Gaussian distributions of $Ar^{9+}$ ions, centred at a depth ~ 75 nm inside the material, shown in inset of figure 2. The estimated penetration depth of 81 KeV $Ar^{9+}$ ions is 100 nm (range ~72 nm and longitudinal struggling ~31 nm) much less as compared to the sample thickness (~500 nm) measured by RBS technique. Hence the implantation generated defects viz. vacancies and interstitials are extended from the surface to 100 nm depth, i.e. in the subsurface region (few atomic layers below the film surface) according to SRIM simulation results.

Figure 3 shows the XRD patterns of implanted and un-implanted $Zn_{0.95}Mn_{0.05}O$ films. Glancing angle XRD patterns indicate single phase, hexagonal (wurtzite) structure of ZnO films without any trace of secondary or impurity phase(s).

The major diffraction peaks (001), (002), (101) of wurtzite ZnO structure have been observed for all the films. The intensity (101) peaks are slightly higher. Close inspection of the XRD pattern reveals that crystalline quality improves when the film implanted with the intermediate dose and (101) peak becomes more prominent than other two major diffraction peaks. At the high dose intensity of the major three (001), (002), (101) diffraction peaks reduces with respect to background indicating degradation of crystalline quality. But no significant amorphization for this film in the crystalline structure was observed. The system maintains its original structure. Figure 4 shows the two dimensional (2D) AFM and MFM images of un-implanted and implanted with high dose $Zn_{0.95}Mn_{0.05}O$ films. For the un-implanted film the value of grain size is ~ 61 nm, estimated from AFM micrographs. It reduces to ~ 26 nm for the film implanted with low dose and increases to ~ 35 nm for the film implanted with the high dose. A tendency of agglomeration was observed at the high implantation dose, as depicted in figure 4(b). Hence the estimated value of the grain size is larger compared to the sample implanted with low dose. This is consistent with the XRD results also, as evidenced by the lowering in full width at half maxima (FWHM) value of the (101) peak of film implanted with the high dose. AFM micrographs showed that the films are composed of closely packed grains of granular in nature; however a distribution in grain size has been clearly observed in the AFM micrographs. Figure 4 (c) and (d) depicts the MFM images of un-implanted and high dose implanted films respectively. Uniform brightness contrast indicates absence of magnetic impurity particles/clusters. This observation strongly supports the view point of single phase nature of films as seen in the XRD spectra.[13]

The estimated penetration depth (~100 nm) of 81 KeV $Ar^{9+}$ ions is much less than the penetration depth of X-ray inside the material. Therefore XRD peak intensity and FWHM gives information of the overall defective nature (weighted average of contributions from the defective and less-defective regions) within the penetration depth of the X-ray.[16] The XRD pattern of implanted film with high dose shows that FWHM of the (101) peak is lowered along with peak intensity.[9] Due to collision with $Ar^{9+}$ ions several defects, defects complexes have been generated in the material. Majority of such defects/vacancies dynamically anneal. [16-17] A ($S_e$) induced recovery, [16] in ZnO also helps in annealing out of some defects inside the system. This dynamic recovery of implantation generated defects or defect complexes, in ZnO based systems have been reported.[16, 19] Actually this is the reason for the radiation hardness of ZnO and that is happening here also.

Optical study of all $Zn_{0.95}Mn_{0.05}O$ thin films has been performed by UV-Visible spectroscopy. Inset of figure 5 indicated high transmittance (~85%) for all the films in the visible region and shows sharp fundamental absorption edge around 380 nm. High transparency of the films enhances their potential for practical application as transparent electrodes. The optical absorption coefficient 'α' is defined as:

$$I = I_0 \exp(-\alpha d) \quad \text{and} \quad \alpha = (1/d) \ln(1/T) \; \ldots\ldots\ldots\ldots\ldots\ldots \; [1]$$

Where, I is the intensity of transmitted light, $I_0$ the intensity of incident light and d the film thickness. And transmittance (T) is defined as $I/I_0$. With the knowledge of the thickness (d) of the films and transmittance T, α was calculated by using the above expression. In case of direct transition, the absorption coefficient can be expressed as:

$$(\alpha h\nu) = A(h\nu - Eg)^{m/2} \; \ldots\ldots\ldots\ldots\ldots\ldots.. \qquad [2]$$

Where A is a constant, hν is the photon energy, Eg is the optical band gap and m is a constant that depends on the nature of semiconductors, m = 1 for a direct transition, whereas for indirect band gap semiconductor m = 4. Figure 5 shows variation of $(\alpha h\nu)^2$ against hν for the un implanted and implanted films. The linear portion of the $(\alpha h\nu)^2$ against hν plot is extrapolated to intersect the energy axis, as Eg = hν (for direct band gap) for $(\alpha h\nu)^2 = 0$. To observe the effect of doping in band gap the plot of $(\alpha h\nu)^2$ against hν of un-doped ZnO film has also been shown. The estimated band gap of ZnO film is 3.25 eV which is less than the reported value 3.37eV. [19-20] The shift in band edge to the lower energy side has been observed for $Zn_{0.95}Mn_{0.05}O$ films [21] and shown in figure 5. The estimated band gap of $Zn_{0.95}Mn_{0.05}O$ film is 3.06eV. The decrease in band gap with Mn doping has already been observed in case of bulk system[6] as well as thin films. [13] It was attributed to the periodic variations in potential within the grain due to trapping of impurities with doping.[6, 13] In the absorption spectra any significant change in band edge of the implanted $Zn_{0.95}Mn_{0.05}O$ films in comparison with un-implanted film hasn't been found. It also implies that significant amorphization in the crystalline structure hasn't taken place. Generally significant amorphization may leads to reduction in band gap. [22] This is also in conformity with the XRD results. Prominent absorption band just below the absorption edge was observed (~2.8 to 3 eV) for the $Zn_{0.95}Mn_{0.05}O$ films. This type of mid gap absorption has been reported in ellipsometry measurements of $Zn_{1-x}Mn_xO$ thin films[23] as well as in the optical transmission spectra of $Zn_{1-x}Mn_xO$ films.[24] The implanted films (particularly the film implanted with high dose) indicate that absorption at low energy region i.e. below the band edge increases as shown in figure 6. Increased absorption in the visible region appears due to transitions between the intra-gap levels related to some defects such as $V_O$, $I_{Zn}$, or $V_{Zn}$ present in the system. Similar kind of defect induced visible absorption has been observed by Agarwal *et al.* [22] in the optical absorption spectra of ZnO thin films after irradiation with 100 MeV $Au^{8+}$ ions.

The effect of disorder due to implantation is more pronounced in band tail region of the absorption spectra. Significant band tailing with implantation has been observed as shown in figure 6. Iribarren *et al.* [25] pointed out that band tail parameter represent the bulk-defect as well as grain-boundary trap concentrations in polycrystalline semiconductors. Actually band tail parameter ($E_0$) reflects the overall defects concentration in samples. [6, 26] Absorption coefficient below the band edge ($E < Eg$) should vary exponentially with absorbed photon energy (E).

$$\alpha (E) = \alpha_0 \exp(E/E_0) \ldots\ldots\ldots\ldots\ldots \quad [3]$$

where $\alpha_0$ is a constant and $E_0$ is the band tail parameter. $E_0$ can be estimated from the reciprocal of the slope of the linear part of the $\ln(\alpha)$ versus E curve ($E < Eg$). The variations of $E_0$ with respect to doses of implantation have been presented in the inset of figure 6. $E_0$ increases dramatically upon ion implantation. The trend of variation of $E_0$ from low dose to high dose implanted films is slightly decreasing. Generally ion implantation generates different types of vacancy/vacancy clusters such as $V_O$, $V_{Zn}$, Zn antisite ($Zn_O$), Oxygen antisite ($O_{Zn}$) in the system.[16, 19] It was reported that presence of vacancy or vacancy clusters in disordered ZnO affects the optical absorption process in two separate ways, close to the band edge.[26] Sufficient number of $V_O$ related defects in the system causes shifting in the band edge (formation of deep centers) and $V_{Zn}$ or their complexes cause tailing of the valence band. [26] When both type defects are presents then their relative effects on band tailing are superimposed. According to Toumisto *et al.*[27] substitution of Mn at the Zn site suppresses the formation of $V_O$ type defects. Also Xu *et al.* [8] demonstrated that donor traps are suppressed with Mn doping however deep acceptor states has been generated with doping in the system for Mn doped ZnO thin films. The deep acceptors have been identified as $V_{Zn}$. In one of our recent work, reduction in visible emission intensity centred ~ 525 nm (assigned as due to $V_O$ type defects) with Mn doping in ZnO has been found. [13] It also indicates that $V_O$ type defects quenches in the system with Mn doping. An increasing value of $E_0$ with higher milling time was reported [6] for Mn doped ZnO bulk system and the value of $E_0$ has been assigned to the representative of acceptor defects mostly $V_{Zn}$ or $Mn_{Zn}$ present in the samples. So $E_0$ might qualitatively represent acceptor type defects for Mn doped ZnO system. However $Ar^{9+}$ implantation in $Zn_{0.95}Mn_{0.05}O$ films can also creates $V_O$ types defects in the system. Hence PL spectroscopy may be useful for further understanding the nature of defects present inside the system.

Figure 7 represents PL spectra of un-implanted and implanted $Zn_{0.95}Mn_{0.05}O$ films. In the PL spectra of the films it has been observed that near band edge excitonic (NBE) emission (~380 nm) becomes almost quenches.[13]This may be due to generation of some non-radiative defect centers inside the system arising out of doping. Similar reduction effects of luminescence were reported in Mn and Co doped ZnO nanorods.[28] Possibly doped cations provide competitive pathways for recombination, which results in quenching of the emission intensity.[13] We have reported earlier that the overall emission intensity, NBE emission (375-385nm) as well as visible emission in the wavelength range (450-550nm) reduces considerably with increasing doping concentration.[13] The luminescence in higher wavelength (400-550nm) region as shown in figure 7 possibly originates from intrinsic defects in the films. The origin of intrinsic defects related with visible emission has not been fully understood due to complexity of the microstructure of ZnO.[16, 26, 29] Zhan *et al.* [29] pointed out that visible emissions near blue, blue-green, green-yellow, and red-near infrared regions were due to presence of zinc interstitial ($I_{Zn}$), $V_{Zn}$, $V_O$ and oxygen interstitial ($O_i$), respectively. Three emission peaks centred ~ 418, 441 and 525 nm was observed in the room temperature PL spectra. Violet emission observed at ~ 418 nm (2.96 eV) is due to electronic transition from a shallow donor energy level of neutral $I_{Zn}$ atoms to the top level of the valence band. Blue emission centred on ~ 441 nm (2.79 eV) is due to singly ionized $V_{Zn}^{-}$.[30] Green emission centred at ~525 nm (2.37 eV) is probably caused by $V_O$ defects introduced in thin films.[13, 29] $V_O$ can exist in three different charge states in the ZnO lattice as follows: $F^0$ (doubly occupied), $F^+$ (singly occupied), and $F^{2+}$ (unoccupied) respectively. The emissions peak observed ~525 nm in the PL spectra is due to $F^+$ vacancy types defects.[13, 29]Deep level emission is controlled by the types and concentration of the corresponding defects. The intensity of the visible emission peak centred ~ 525 nm for implanted film with low dose increases in comparison to un-implanted film. However with increasing dose of implantation it decreases. Ultimately for the high dose the intensity of visible emission ~ 525 nm is almost comparable with the un-implanted film. The increase of green emission intensity at the low dose can be thought of as the enhancement of $V_O$ concentration. At the high dose regime reduction in visible emission intensity ~ 525 nm indicates saturation of defect concentration in the system.[16-17]The intensity of another visible emission centre ~ 418 nm has found to be increased with increasing ion implantation; except for the exposure on high dose, upon which it decreases. It is indicative of increasing $I_{Zn}$ concentration in the films with increasing dose of implantation. Finally with high dose of implantation the concentration of $I_{Zn}$ decreases. The intensity of another visible emission centre ~ 441 nm has been found to increase at the intermediate dose and finally reduced with further increase of implantation dose. Overall the fact can be interpreted as with increasing dose, the concentration of $V_{Zn}$ deceases. However the changing pattern of defect concentration of $V_O$, $V_{Zn}$ and $I_{Zn}$ with increasing implantation dose cannot follow any unique pattern. A competition between defect generation and recovery (ionization induced recovery contributes to some extent) for different varieties of defects leads to some kind of complex formation and destruction routes.

The M-H variations at 300K, 250K and 200K of un-implanted and implanted (with different dose) $Zn_{0.95}Mn_{0.05}O$ films have been presented in figures 8 (a), 8(b) and 8(c) respectively. The saturation magnetization of the film was determined by subtracting the diamagnetic contribution of the substrate from the experimental raw data using the following equation:

$$M_S(H) = M_{Exp}(H) - \chi H \quad \ldots\ldots\ldots\ldots\ldots\ldots\ldots \quad [4]$$

Where $M_S$ (H) is the saturation magnetization of the films, $M_{Exp}(H)$ is the experimental M-H data and $\chi$ is diamagnetic susceptibility of the substrate. The well known Arrott plot for all the films at 300K is shown in inset of figure 8 (a). The plot shows convex curvature, confirming thereby the presence of spontaneous magnetization, i.e., the FM ordering of the samples.[31] The values of saturation magnetization ($M_S$) and coercive field ($H_C$) are presented in table 1. The increasing tendency of $M_S$ and remanent magnetization ($M_R$) with decreasing temperature for all the films is quite expected. At 300K, the values of $M_S$ and $M_R$ deceases for implanted film with low dose in comparison to un-implanted film. For the film implanted at intermediate dose the value of $M_S$ and $M_R$ increases a bit but still lower than un-implanted film. Finally for the implanted film with high dose the value of $M_S$ and $M_R$ is highest. The highest value of $M_S$ is 0.69emu/gm for the implanted film with high dose. This value is low as compared to theoretically predicted value (5$\mu_B$/Mn atom). Primarily the reason attributed for such low values of magnetization is the presence of AFM coupling between neighboring Mn atoms. [4-6, 13] However the value is quite high so far as reported [6, 10, 13] values for Mn doped ZnO system are concerned. The fact of observing comparatively higher value of saturation magnetization for Mn doped ZnO thin film in comparison to its bulk counterpart seems to be noteworthy from the point of spintronic based device application of DMS. It has been observed that $H_C$ increases with low dose of implantation, but decreases significantly at the intermediate dose and finally at high dose of implantation it increases again and attains the maximum value. However Hc decreases (except for the film implanted with low dose) at lower temperature. This type of anomalous temperature dependence of Hc has been reported earlier.[32] It may be ascribed to the anomalous temperature dependence of the magneto-crystalline anisotropy originating from a special microstructure.[33] The nature of variation is more or less similar to reported observation in case of chemically synthesized $Fe_3O_4$/ZnO nanocmposites.[33] The nature of variation of $M_S$ and $M_R$ at 250K and 200K are remarkably similar with that of at 300K signifies a steady and uniform magnetic properties of all the films. The overall magnetic property has been upgraded with ion implantation if un-implanted and high dose implanted films are compared; the optimized dose of $Ar^{9+}$ ion implantation for strong ferromagnetic Mn doped ZnO film is assigned as $1x10^{16}$ ions/cm$^2$. Further the film implanted with high dose has shown 85% transparency in the entire visible wavelength spectrum [inset of figure 5] indicating its wide applicability as transparent electrodes. So the application potential of the film implanted with dose of $1x10^{16}$ $Ar^{9+}$ ions/cm$^2$ seems to quite high in both spin and transparent electronics. The inset of figure 8(c) indicates M-H variation for un-implanted and high dose implanted films at 325K. The value of $M_S$ and $H_C$ are presented in table 1. The value of $M_S$ (0.76 emu/gm) for high dose implanted film at 325K is higher as compared to un-implanted film. This fact is quite obvious considering the values of $M_S$ at 200K, 250K and 300K. The values of $M_S$ are higher in comparison to the values at 300K for both films. Hence above RT both the films exhibit increasing magnetization with elevated temperature. The observation is quite unexpected but it is not a new observation, similar trend for Mn doped ZnO films was reported.[34] It is indicative of spontaneous cooperative magnetization.[34] The values of $H_C$ are also quite appreciable for both the films and high dose implanted film shows quite high value. In comparison to $H_C$ values at 300K un-implanted film shows nearly same value whereas for high dose implanted film $H_C$ decreases a little bit. Hence the M-H variations for the two films indicate that ferromagnetic transition temperature is well above 325K for this system. This fact seems to be quite interesting so far as its utilization in spin electronic device is concerned as discussed in the section of introduction.

The M-T variations of all the films under ZFC and FC condition at 150Oe magnetic field have been presented in figure 9. Identical plots in the entire measuring temperature range (15 to 300K) of the M-T variations under ZFC and FC conditions eliminates the possibility of presence of superparamagnetic / spin glass behavior. [35] A rapid decrease in magnetization with increasing temperature below 50K was observed. Above 50K, the magnetization becomes much less sensitive to temperature, and is sustained up to 300K. This kind of temperature dependence of magnetization suggests coexistence of paramagnetic component, dominant at low temperatures and a ferromagnetic component with a $T_C$ higher than 300K.[7, 34] The existence of paramagnetic interaction is prominent at low temperature and almost thermally independent behavior of magnetization M(T) above 50K. [21, 32] It indicates existence of paramagnetic components at low temperature and ferromagnetic components with Tc greater than 300K in the system.[7] ZFC & FC magnetization curves (M-T) although follow the similar trend throughout the measured temperature range but the difference in magnitude of ($\Delta M = M_{FC} - M_{ZFC}$) is non zero up to 300K [as presented in the inset of figure 9 (a) and (d) for un-implanted and high dose implanted films]. The overall value ΔM has been found to decrease with increase in temperature as expected. The subtraction of $M_{ZFC}$ from $M_{FC}$ data eliminates paramagnetic and diamagnetic contributions. Simultaneously, a nonzero difference up to 300K further confirms the presence RT ferromagnetic ordering in the films.[13, 35]Notably, smoothness of $M_{FC}$ (T) and $M_{ZFC}$ (T) curves throughout the entire temperature range from 15K to 300K confirms the absence of any segregation of any secondary/impurity phase in the films. Hence the observed FM in all of the films is intrinsic, and which is the very basic objective of any DMS.

The observed magnetic properties of un-implanted film and all of the implanted $Zn_{0.95}Mn_{0.05}O$ films seem to quite correlate with intrinsic defects present in the system. The optical characterization by UV-Visible spectroscopy indicates generations of some stable defects inside the system because strong absorption has been observed below the band edge, shown in figure 6. However strong and direct correlation has been noticed by considering the PL spectra (figure 7). The blue emission peak ~ 441 nm and the green emission peak ~ 525 nm have been assigned as due to singly ionized $V_{Zn}$ and singly occupied $V_O$ respectively. The variation of ratio of intensity of peaks ($I_{blue\ emission}/I_{green\ emission}$) has been plotted against ion implantation dose in figure 10. Further the variation of $M_S$ has also been plotted against ion implantation dose in figure 10. The close resemblance of variation of intensity ratio ($I_{blue\ emission}/I_{green\ emission}$) and $M_S$ against ion implantation dose is the most striking feature of this article. Therefore increase of concentration of singly ionized $V_{Zn}$ and/or decrease in concentration of singly ionized $V_O$ is responsible for increasing magnetization of this system. Alternately, $V_{Zn}$ increases ferromagnetic ordering whereas the role of $V_O$ may be opposite. So a delicate balance between the concentration of $V_{Zn}$ and $V_O$ is required for tuning the ferromagnetic properties of Mn doped ZnO films. It confirms the important role played by lattice point defects in understanding and tailoring the magnetic properties of the Mn doped ZnO film system. Further FM observed in nanocrystalline ZnO[10] or in single crystalline ZnO after Ar implantation[37] is of much lower magnitude (~ $10^{-3}$ emu/g) as compared to the observation of this work. Thus it appears that defects ($V_{Zn}$) and substituted Mn ions at the Zn site both[6, 10] are the important ingredients for developing ferromagnetic coupling in DMS system.

A remarkable microscopic mechanism was proposed involving formation of BMP upon introduction of p-type dopants. [2, 38]The mechanism is as follows: magnetic interactions between defect bound valence band holes and $Mn^{2+}$ ions align $Mn^{2+}$ spins with respect to one another, forming a BMP. With increase in defect concentration different overlapping BMPs form an extended ferromagnetic domain. In this work, the FM arises from similar type of mechanism. Apart from $Mn^{2+}$ ions any further doping element hasn't been utilized. Singly ionized $V_{Zn}$ and singly occupied $V_O$ plays the role of acceptor and donor of electrons respectively. The activation of acceptor states is necessary in order to achieve FM. The carriers arising from singly ionized $V_{Zn}$ type defect may be delocalized, but with low mobility, thus yielding low conductivity. The BMP formation has been manifested by the presence of singly ionized $V_{Zn}$ and $Mn^{2+}$ ions. The singly ionized $V_{Zn}$ interacts with d-electrons of $Mn^{2+}$ ions. The loosely bound extra electron of singly occupied $V_O$ may be trapped at the singly ionized $V_{Zn}$ site. It is compensating the interaction of d-electrons of $Mn^{2+}$ ions with singly ionized $V_{Zn}$. The trapping process hinders the formation of BMP. In other words, the addition of donors to the system via singly occupied $V_O$ will move the Fermi energy level up, resulting in a decrease in effective acceptor density and a reduction in magnetization. The findings are consistent with those reported for Mn doped ZnO in which the intrinsic defect-mediated donor states are high in density.[2, 39]

The PL emission peak ~ 441 nm (blue emission) is strongest in intermediate dose implanted film which indicates the presence of $V_{Zn}$ in maximum proportion. The film shows weaker FM than un-implanted and high dose implanted films because simultaneously strong presence of PL emission peak ~ 525 nm (green emission) responsible for presence of $V_O$ in substantial proportion that obstructs BMP formation. In case of high implanted dose film though the PL emission peak ~ 441 nm (blue emission) is comparatively less intense, however the PL emission peak ~ 525 nm (green emission) depressed significantly. Actually the intensity ratio ($I_{blue\ emission}/I_{green\ emission}$) is highest for high dose implanted film and that causes highest generation of BMP and strongest FM among others. So the role of $V_{Zn}$ is positive in achieving FM [8, 10] whereas $V_O$ is playing just opposite role by obstructing the formation of BMP. Hence a tuning for increasing generation of singly ionized $V_{Zn}$ and simultaneous downsizing the generation of singly occupied $V_O$ may develop a system comprising more number of overlapping BMPs i.e. large size of ferromagnetic domain. The proposed mechanism of FM is modified from usual p-type doping induced [hole mediated] BMP formation mechanism.[38] Actually instead of interaction of p-type dopants and d-electrons of $Mn^{2+}$ ions, singly ionized $V_{Zn}$ interacts with d-electrons of $Mn^{2+}$ ions. Another noted point is that conductivity in defect rich Mn doped ZnO samples is essentially low because high compensation of carriers from donor ($V_O$, $I_{Zn}$) and acceptor ($V_{Zn}$).[9] The low conductivity of the films unambiguously rule out the possible carrier mediated exchange, such as RKKY interaction which is based on the exchange coupling between the magnetic ions and free carriers. Further, conventional super exchange interactions cannot produce long-range magnetic order at very low concentrations of magnetic cation.[40-41] The ferromagnetic exchange here is mediated by singly ionized $V_{Zn}$ that form BMPs. These BMPs overlap and hence responsible for long range $Mn^{2+}$ - $Mn^{2+}$ ferromagnetic coupling in Mn doped ZnO. We propose the high Tc (above 325K) in these films may arise from the hybridization and charge transfer from the singly ionized $V_{Zn}$ states to 3d states of Mn ions near the Fermi level. [41]

Inset of figure 10 shows the generation of $V_{Zn}$ and $V_O$ estimated according to SRIM at 81 KeV $Ar^{9+}$ for a film of thickness about 500nm. The figure clearly exhibit that number of $V_{Zn}$ is higher than the number $V_O$ present in the system. However figure 7 indicates that in all films proportion of $V_O$ is higher than the proportion of $V_{Zn}$. This is due to the fact that substantially $V_{Zn}$ may act as non-radiative defect centre.[26] So PL emission efficiency of the peak due to $V_{Zn}$ is to some extent suppressed. It may be the reason for achievement of strong FM in spite strong presence of $V_O$ [compensating agent of BMP formation]. So strong FM in these films supports SRIM estimated observation of higher number of $V_{Zn}$ than $V_O$.

## IV. CONCLUSIONS

To summarize, 5at% Mn doped ZnO films were synthesized by sol-gel technique and $Ar^{9+}$ ion implantation was made on the as-synthesized films with doses of $5 \times 10^{14}$, $1 \times 10^{15}$ and $1 \times 10^{16}$ ions/cm$^2$. The thickness of the un-implanted film has been accurately estimated from RBS measurement. SRIM calculation indicates ion implantation generated defects fall within subsurface region of the films. XRD pattern indicate hexagonal structure of all films. Grain size of the films estimated from AFM measurement and its nature of variation has been explained. UV-visible spectroscopy measurement indicates no significant variation of band gap for the films. However presence of different varieties of defects in the films particularly for implanted ones has been exhibited from this study. PL spectra clearly demonstrate the presence of $V_{Zn}$, $V_O$ and $I_{Zn}$ type intrinsic defects in the films. The variation of proportion of these defects from un-implanted film to implanted films with different doses has also been interpreted by analyzing PL spectra. In M-H measurement all the films exhibit ferromagnetism well above RT. This fact has been further confirmed from M-T measurement. The presence of any impurity phase has been completely ruled out from XRD, MFM and M-T measurements. In comparison to un-implanted film the films implanted with low and intermediate dose indicate poor magnetic property; however the implanted film with high dose is magnetically best among all. A striking point is that variation tendencies of saturation magnetization and intensity ratio ($I_{blue\ emission}/I_{green\ emission}$) of PL peaks closely resemble with each other. This observation is interpreted as $V_{Zn}$ increases ferromagnetic ordering whereas the role of $V_O$ is opposite. The role of defect species $V_{Zn}$ supports the view point of generation of FM by defect induced BMP. $V_O$ acts as compensating agent to $V_{Zn}$ thereby opposing FM indirectly. Hence the FM in $Ar^{9+}$ implanted Mn doped ZnO films may be tuned by proportion of particular defect species. Finally, in spite of high temperature annealing, strong well above RTFM and high transparency of these films are indicative of their potential in transparent spin electronic device.

**ACKNOWLEDGEMENTS**

This work is financially supported by DST-Government of India and CRNN, University of Calcutta, vide project nos: SR/FTP/PS-31/2006 and Conv/002/Nano RAC (2008) respectively. The author S.K.N. is thankful to UGC for providing his fellowship. We acknowledge VECC, Kolkata, Department of Chemical Engineering, University of Calcutta, CRNN, University of Calcutta, to use their low energy ion beam, AFM, SQUID VSM and PL spectroscopy facilities respectively. We also acknowledge Dr. A. Sarkar for his useful discussion in ion beam implantation.

**Table 1:** Magnetic parameters estimated from M-H variation

| *Sample* | *300K* | | *250K* | | *200K* | | *325K* | |
|---|---|---|---|---|---|---|---|---|
| $Zn_{0.95}Mn_{0.05}O$ film | Ms (emu/g) | Hc (Oe) | Ms (emu/g) | Hc (Oe) | Ms (emu/g) | Hc (Oe) | Ms (emu/g) | Hc (Oe) |
| Un-implanted | 0.55 | 48.50 | 0.88 | 39.2 | 0.97 | 39.9 | 0.67 | 49.0 |
| Implanted with low dose | 0.10 | 62.30 | 0.12 | 72.7 | 0.15 | 71.3 | | |
| Implanted with intermediate dose | 0.27 | 34.7 | 0.36 | 25.1 | 0.41 | 26.1 | | |
| Implanted with high dose | 0.69 | 80.0 | 1.02 | 67.5 | 1.04 | 69.3 | 0.76 | 76.0 |

**Figure Caption**

**Figure 1:** Spectrum of RBS measurement for un-implanted $Zn_{0.95}Mn_{0.05}O$ film.

**Figure 2:** (a) Variation of electronic and nuclear energy loss with energy of the incoming ion beam, as calculated from SRIM. The arrow indicates the incoming direction of the $Ar^{9+}$ ion beam. Inset shows ion ranges inside the samples.

**Figure 3:** XRD pattern of un-implanted and implanted with low, intermediate and high dose of $Ar^{9+}$ ions $Zn_{0.95}Mn_{0.05}O$ thin films.

**Figure 4:** AFM images of (a) un-implanted and (b) implanted with high dose $Ar^{9+}$ ions $Zn_{0.95}Mn_{0.05}O$ thin films. MFM images of (c) un-implanted and (d) implanted with high dose $Ar^{9+}$ ions $Zn_{0.95}Mn_{0.05}O$ thin films.

**Figure 5:** The variation of $(\alpha h\nu)^2$ against photon energy ($h\nu$) for all the thin films. Inset shows the transmittance spectra for all the thin films.

**Figure 6:** The variation of ln ($\alpha$) against photon energy ($h\nu$) curves for all the thin films.

**Figure 7:** PL spectra of the un-implanted and implanted with low, intermediate and high dose of $Ar^{9+}$ ions $Zn_{0.95}Mn_{0.05}O$ films.

**Figure 8:** (a) Magnetization against field (M-H) variation at 300K for all the thin films. Inset shows extended portion the saturation behavior for all the thin films. (b) M-H variation at 250K for all the thin films. (c) M-H variation at 200K for all the thin films. Inset shows M-H variation at 325K and 300K for un-implanted thin film.

**Figure 9:** Magnetization against temperature (M-T) variation under ZFC and FC conditions for (a) un-implanted (b) implanted with low dose (c) implanted with intermediate dose (d) implanted with high dose. Inset of (a) and (d) show the difference between the ($M_{FC} - M_{ZFC}$) moments against temperature for un-implanted and implanted with high dose films respectively.

**Figure 10:** The variation of saturation magnetization and PL intensity ratio ($I_{blue\ emission}/I_{green\ emission}$) against dose of implantation of $Ar^{9+}$ ion beam. Inset shows the total damage, vacancies of both $V_{Zn}$ and $V_O$ estimated according to SRIM for 81 KeV $Ar^{9+}$ beam with units number/ion/nm.

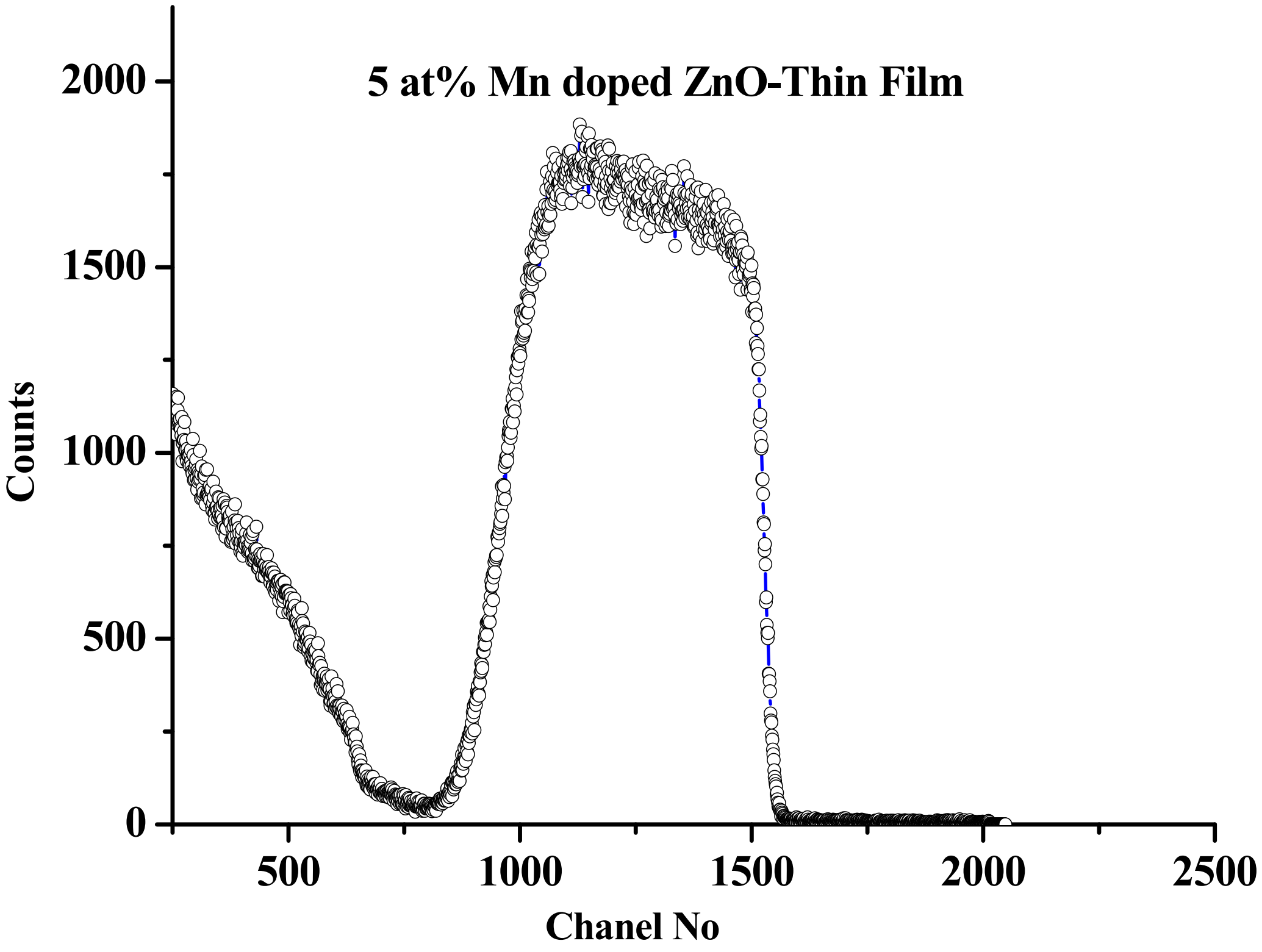

**Figure: 1**

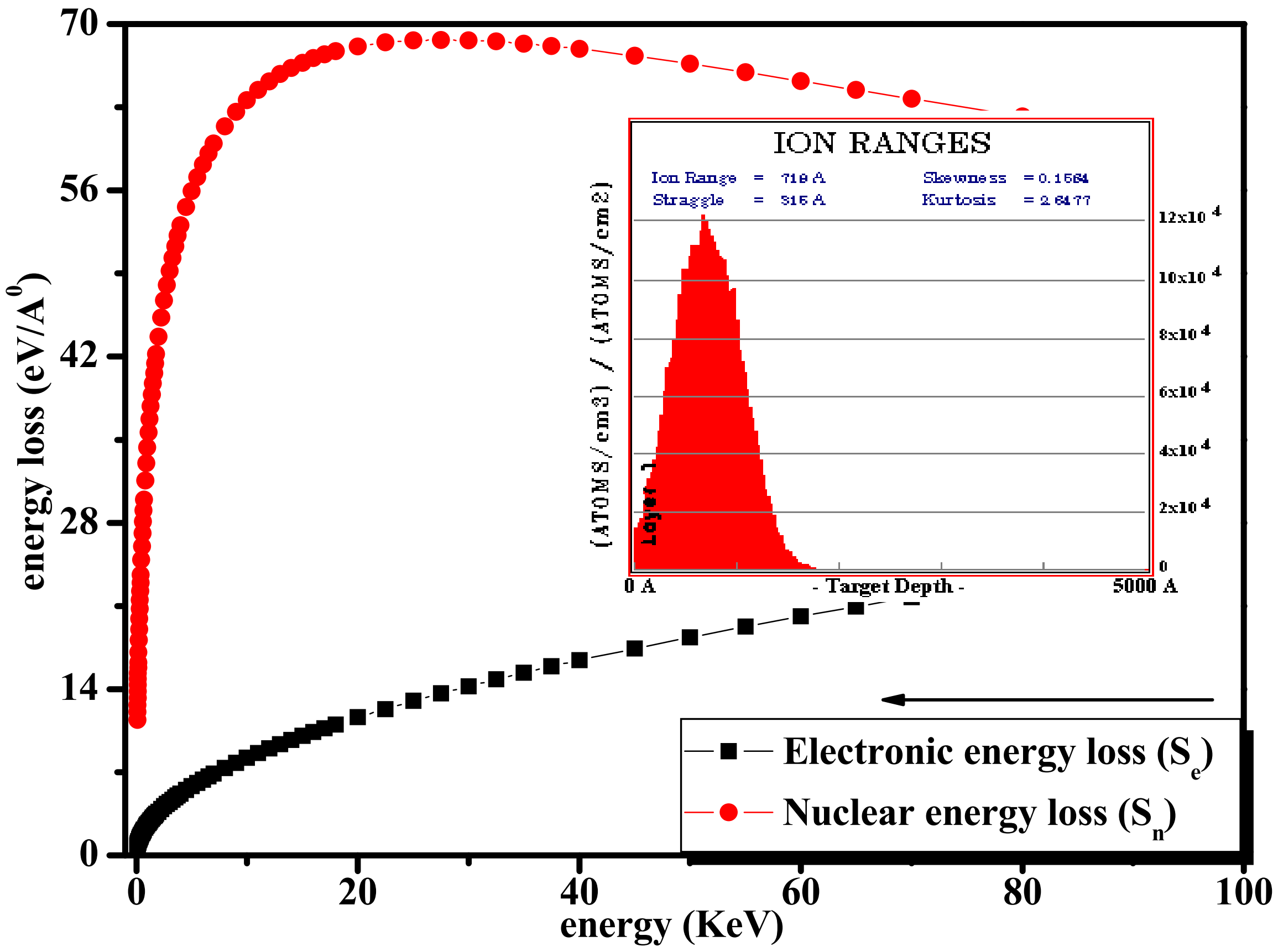

Figure: 2

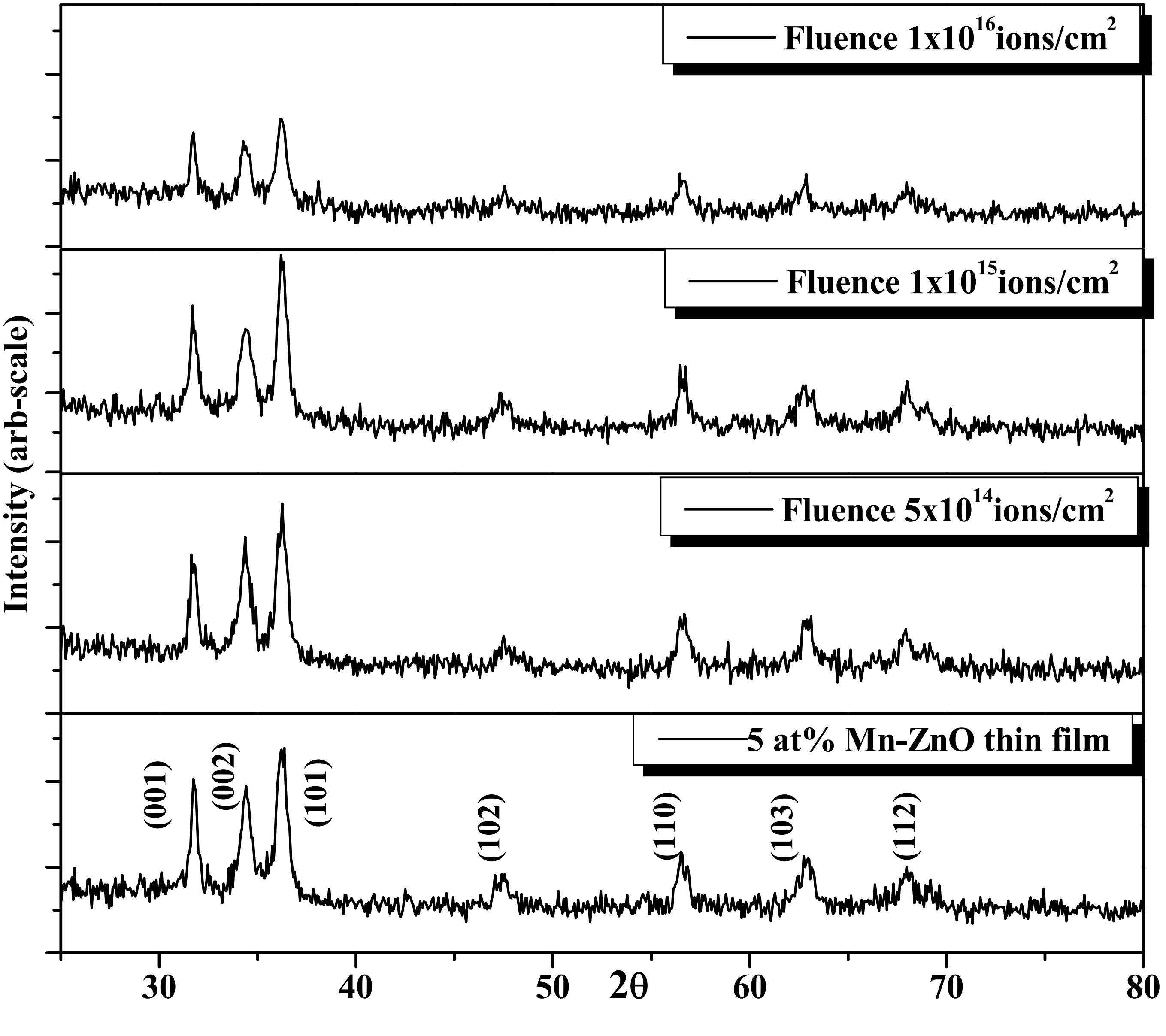

Figure: 3

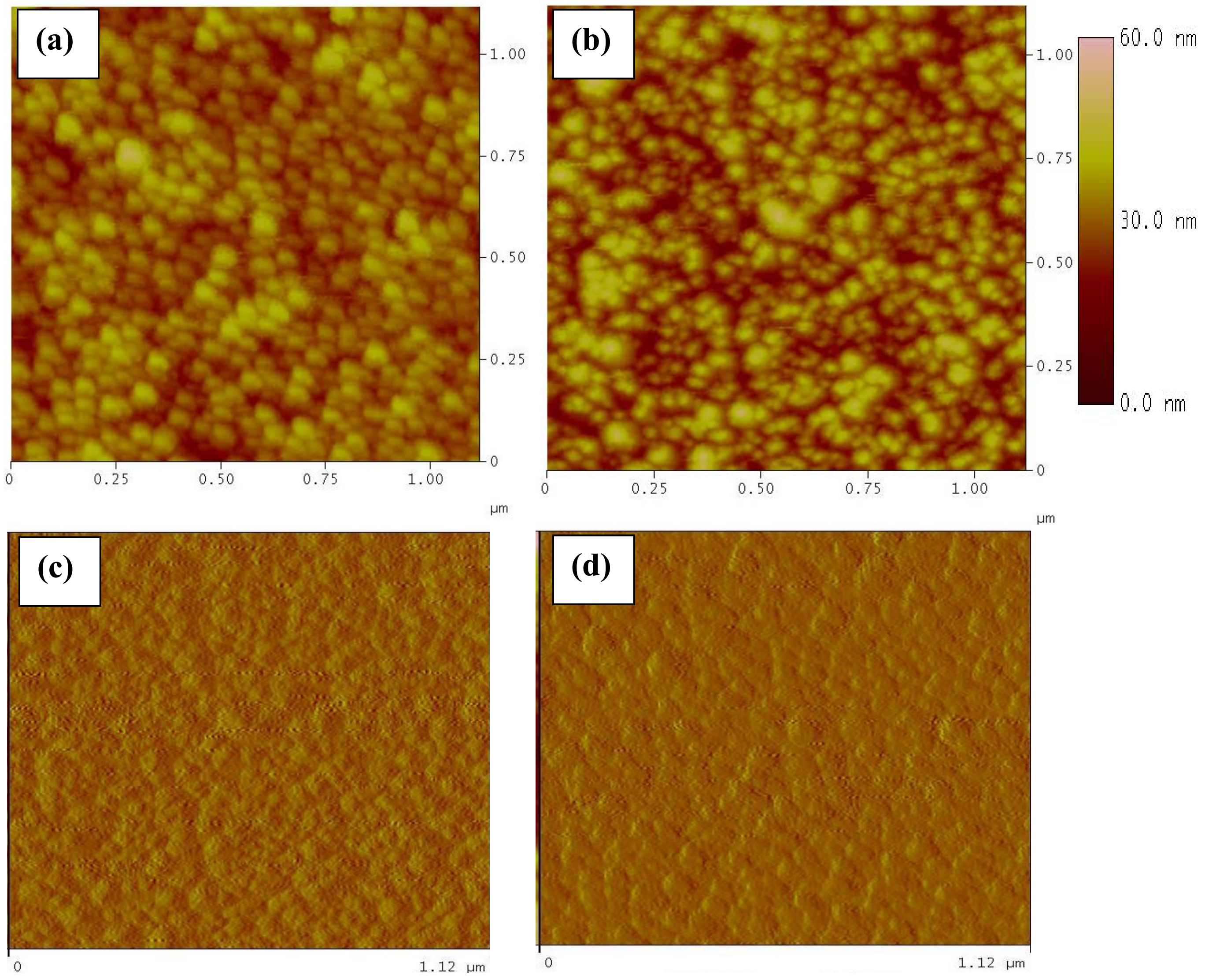

**Figure: 4**

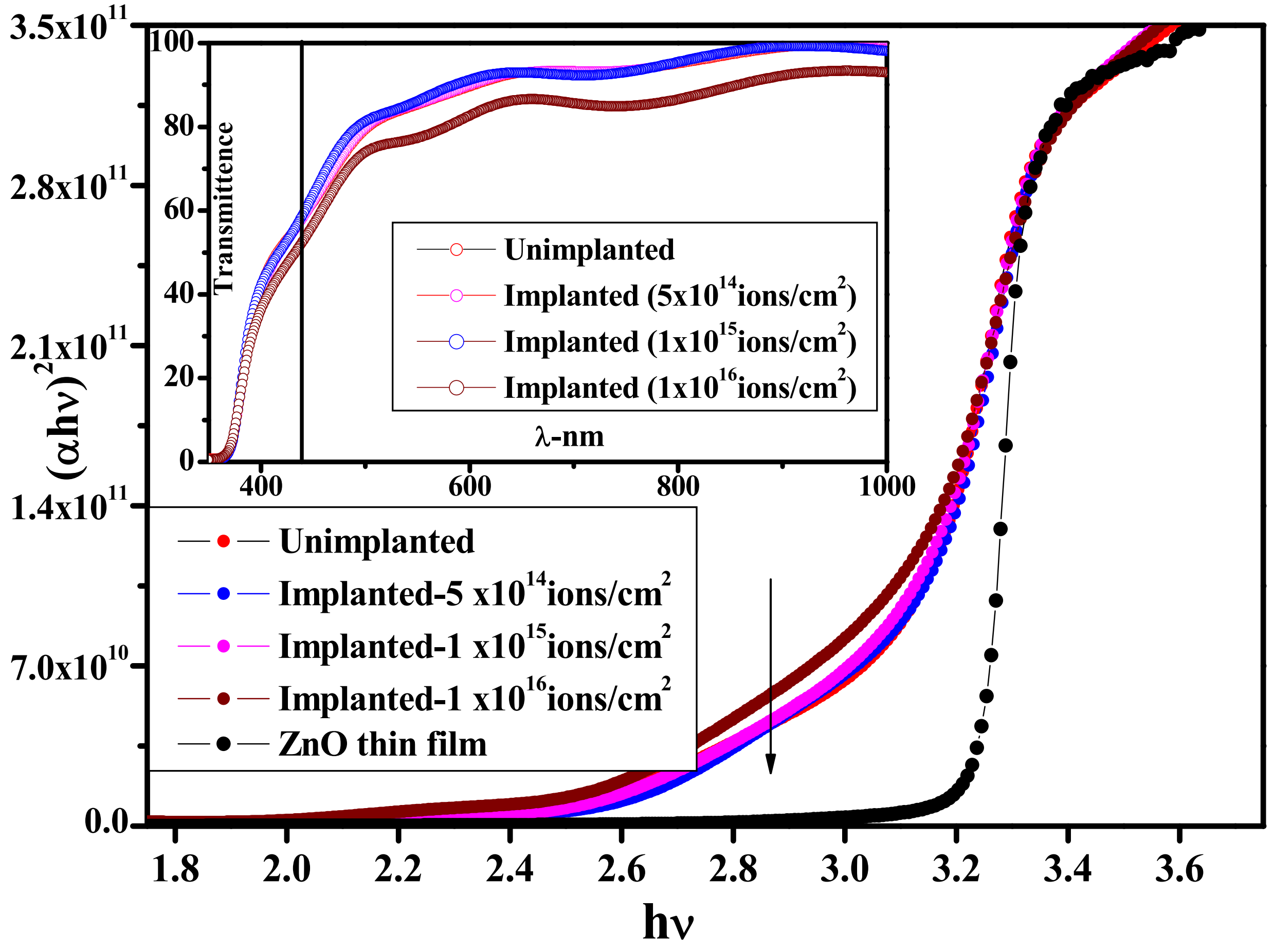

Figure: 5

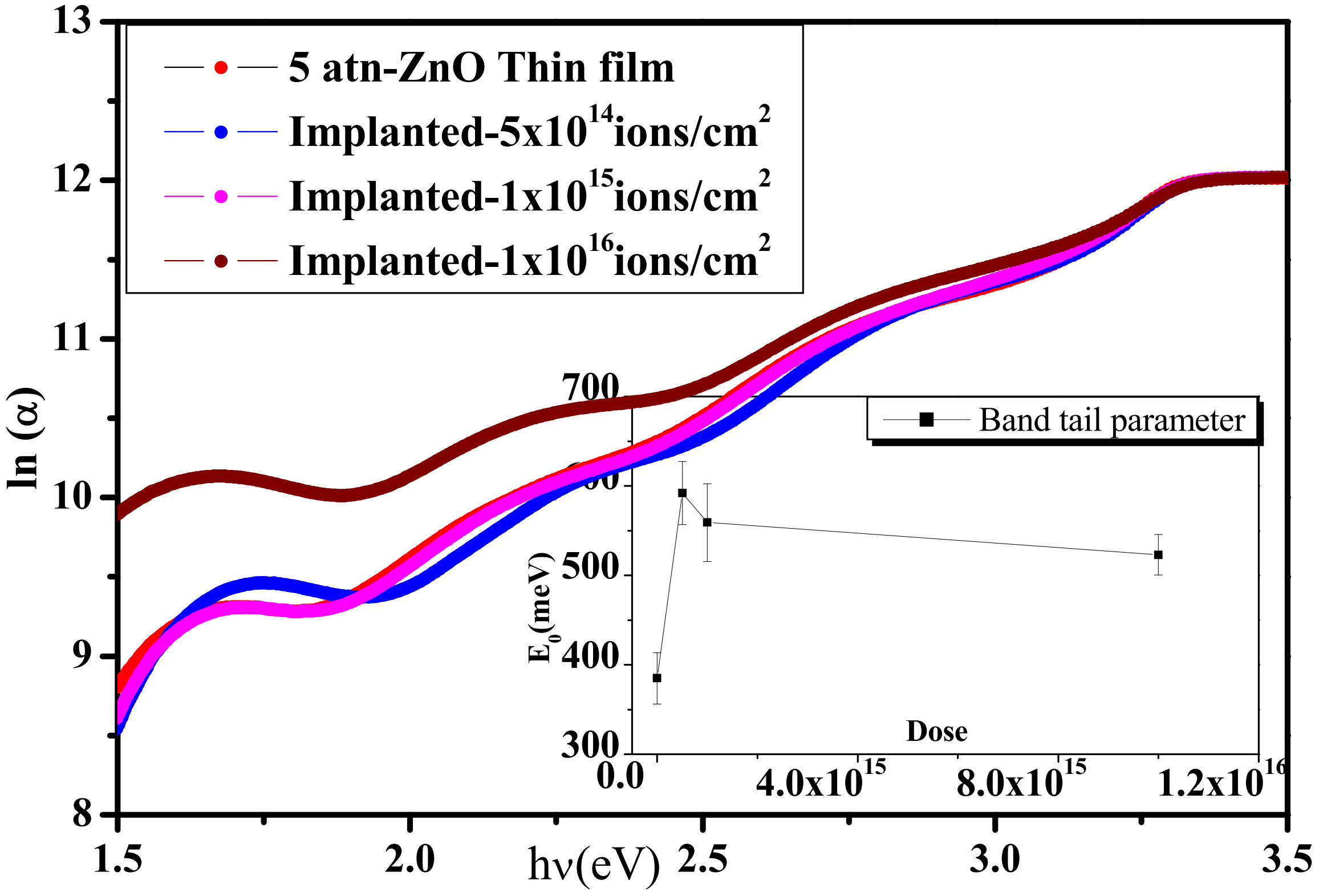

**Figure: 6**

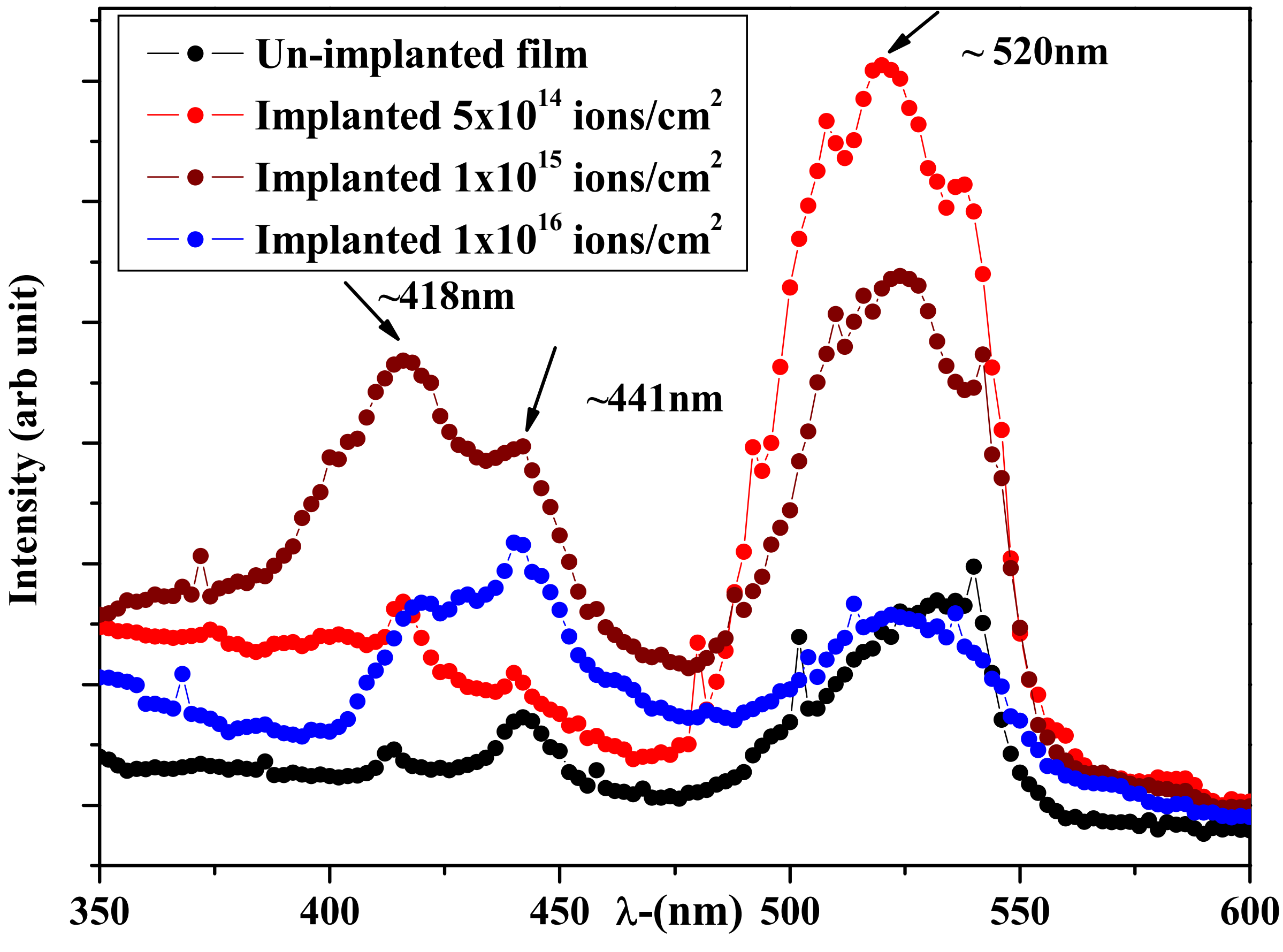

Figure: 7

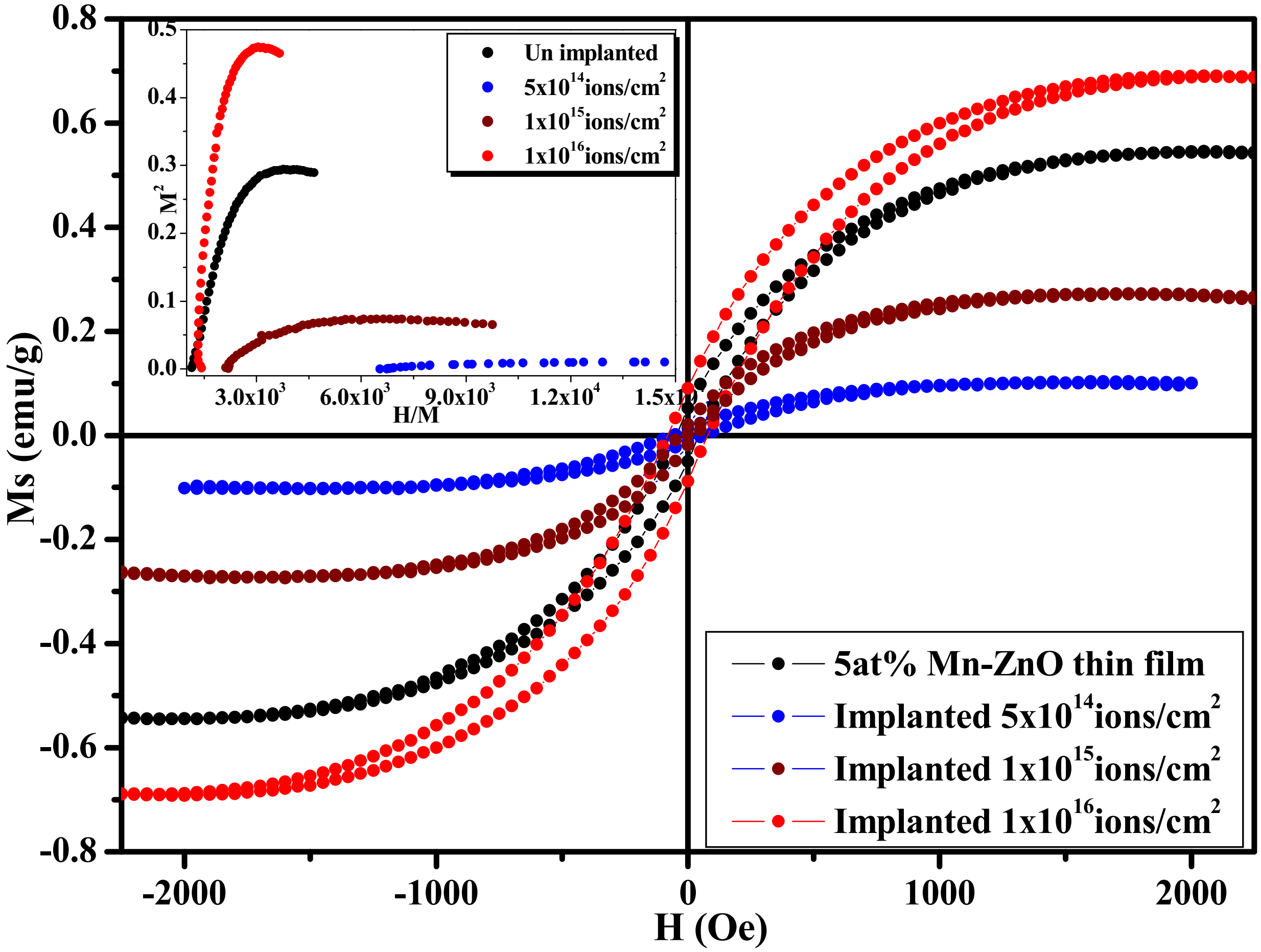

Figure: 8 (a)

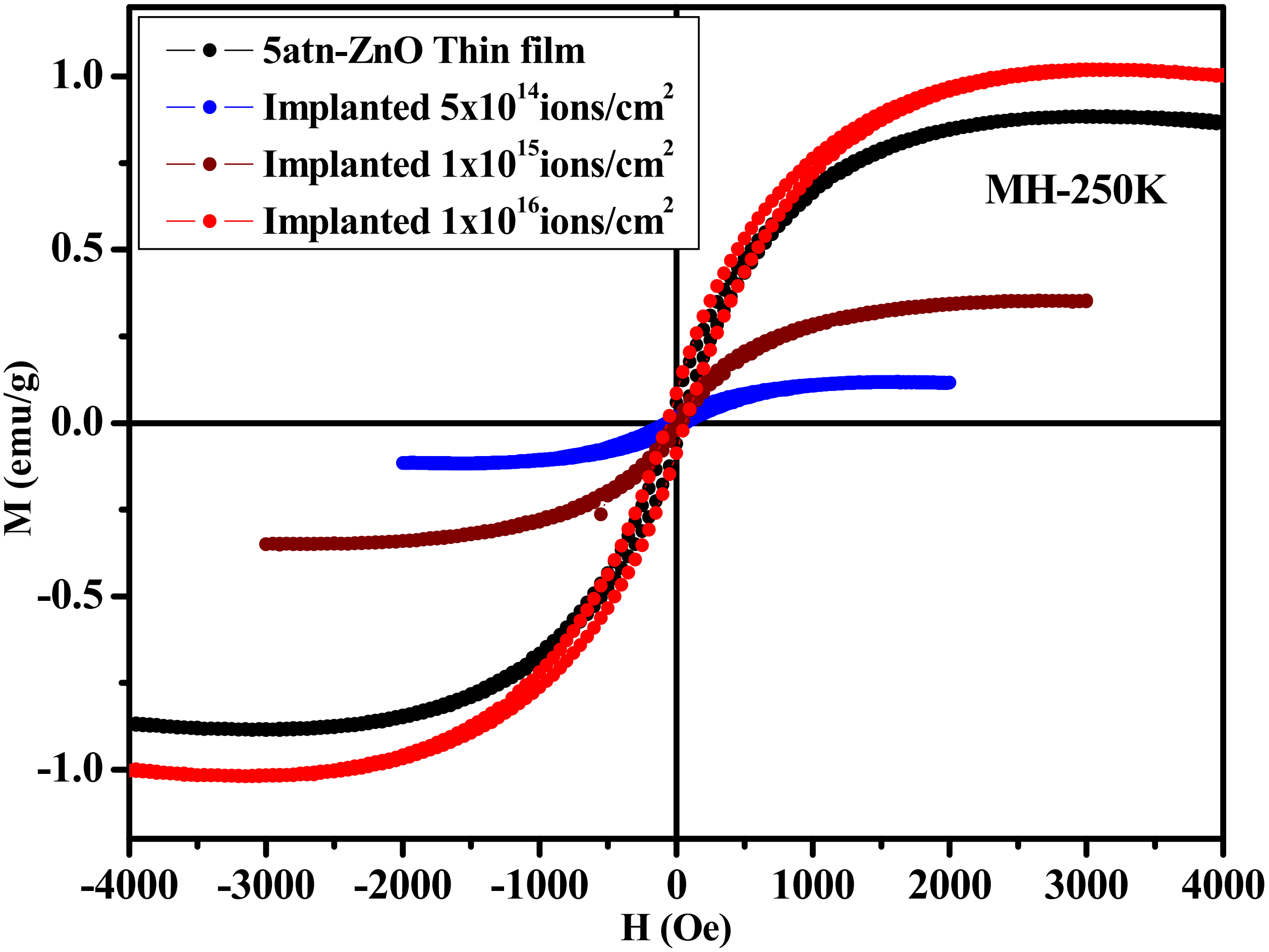

Figure: 8 (b)

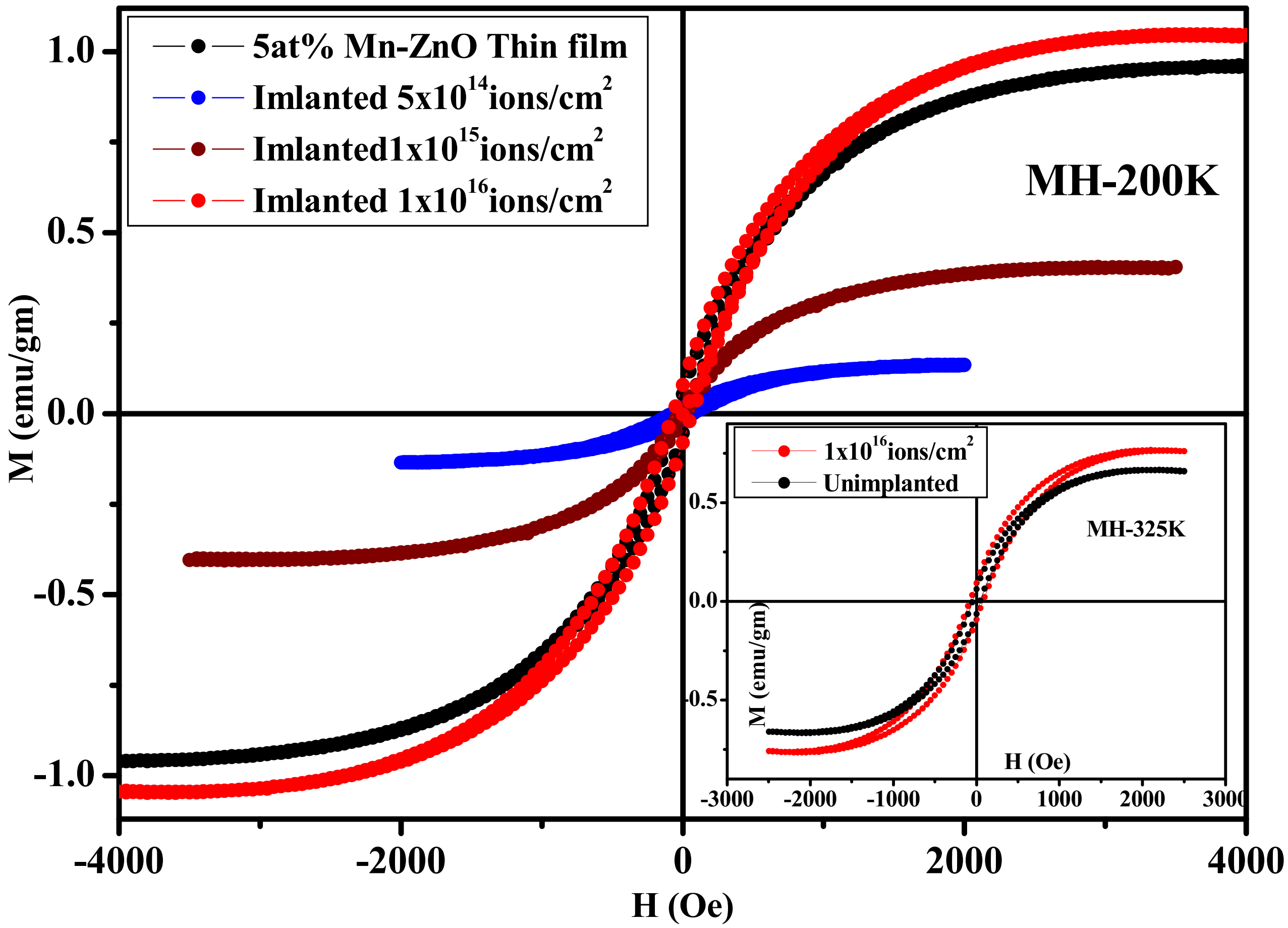

Figure: 8(c)

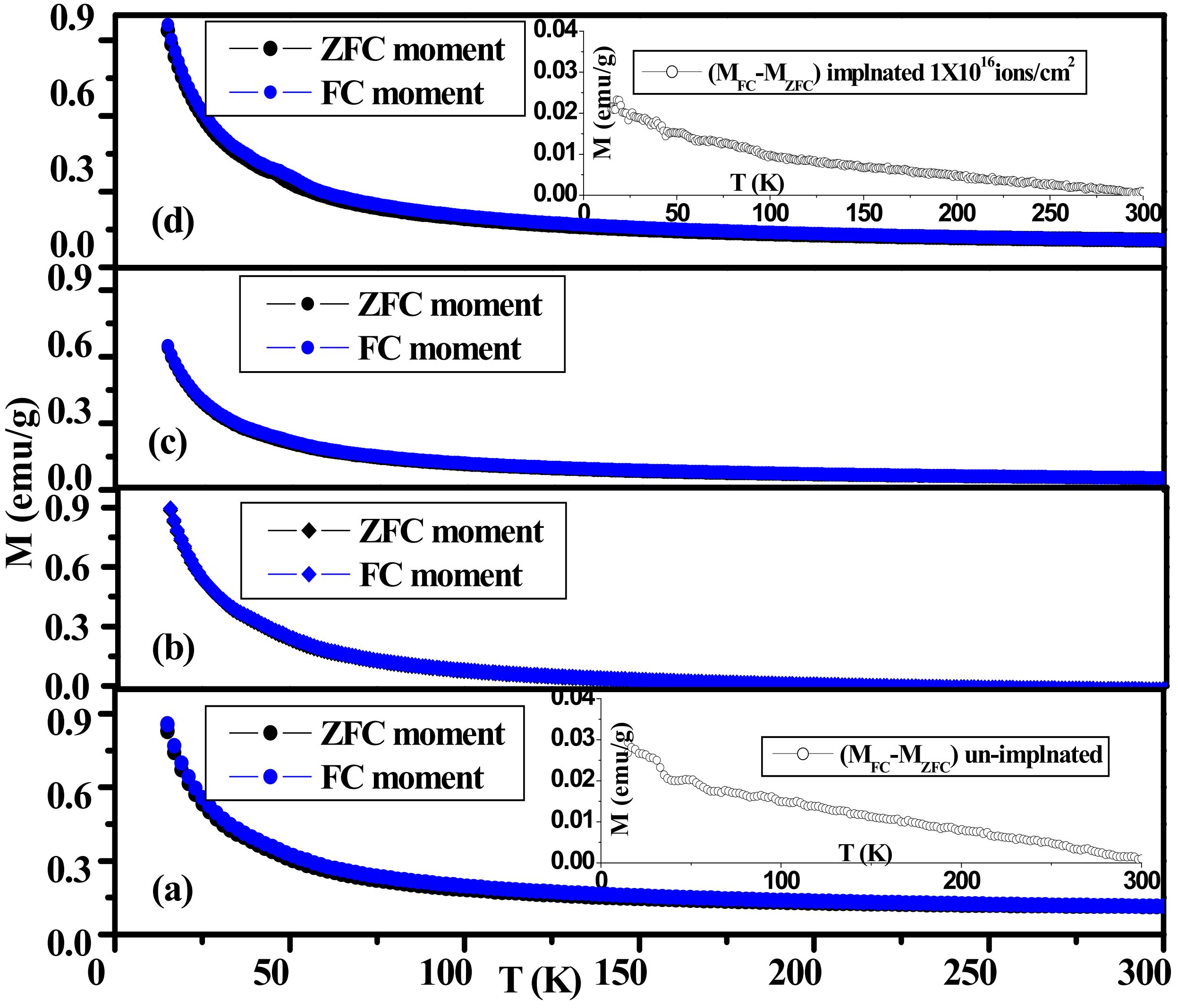

Figure: 9

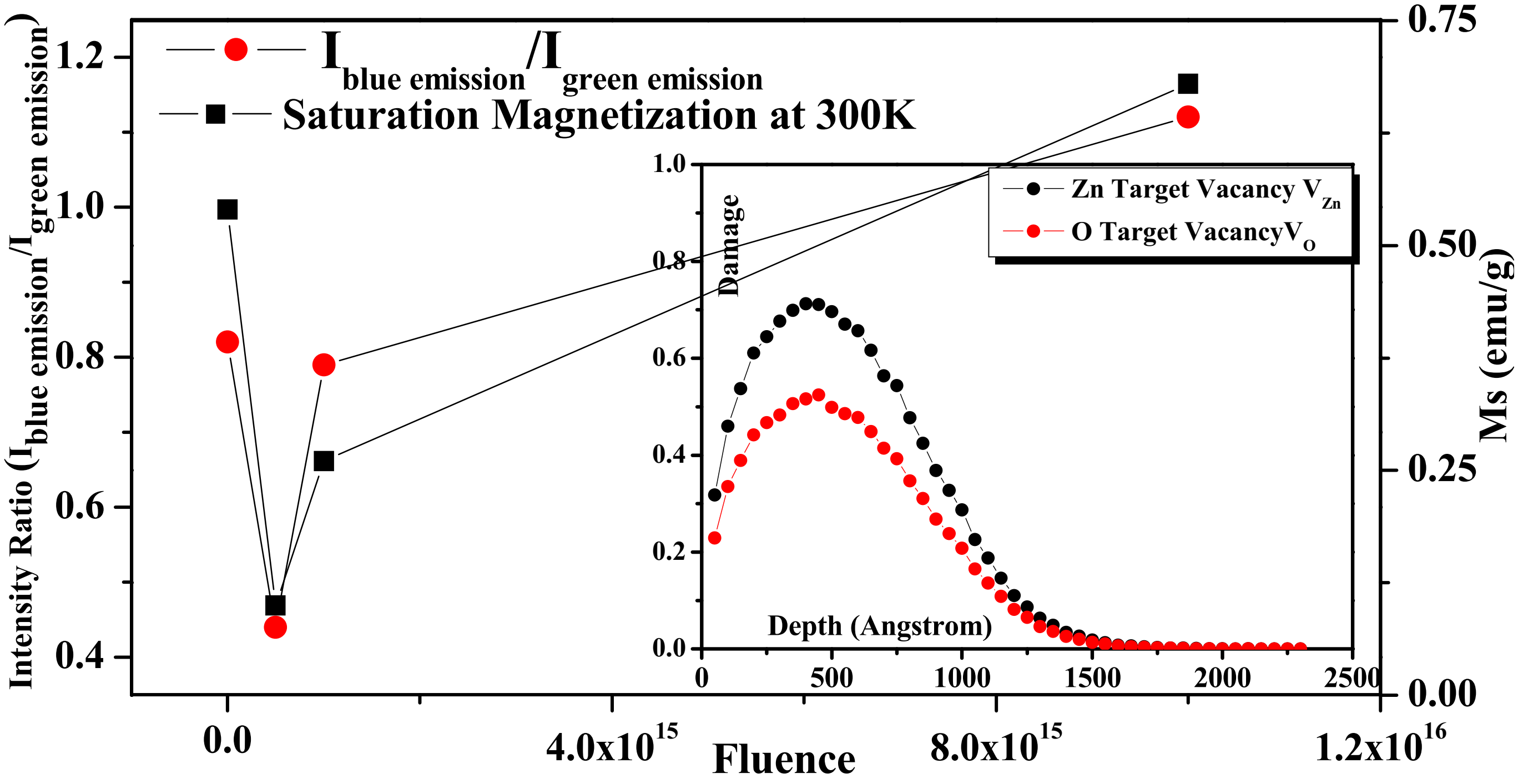

Figure: 10